\documentclass[%
 aps,
 amsmath,amssymb,
 reprint,%
]{revtex4-1}

\usepackage{graphicx}% Include figure files
\usepackage{dcolumn}% Align table columns on decimal point
\usepackage{bm}% bold math
\usepackage[utf8]{inputenc}
\usepackage[T1]{fontenc}
\usepackage{mathptmx}
\usepackage{xcolor}

\begin{document}

%\renewcommand{\theequation}{S\arabic{equation}}
%\renewcommand{\figurename}{Figure S}
%\renewcommand{\tablename}{Table S}
%\renewcommand{\thetable}{\thesection\arabic{table}}
%\titleformat{\section}{\normalfont}{\thesection}{1em}{\MakeUppercase{#1}}
\renewcommand\thesection{\arabic{section}}
\renewcommand{\thesubsection}{\thesection.\arabic{subsection}}
\renewcommand{\thetable}{\arabic{table}}

\title[Sample title]{Machine Learning-Aided First-Principles Calculations of Redox Potentials}
% Force line breaks with \\

\author{Ryosuke Jinnouchi}
\affiliation{ 
Toyota Central Research and Developments Laboratories Inc. %\\This line break forced with \textbackslash\textbackslash
}%
 \email{e1262@mosk.tytlabs.co.jp}
\author{Ferenc Karsai}%
\affiliation{ 
VASP Software GmbH/Sensengasse 8, 1090 Vienna, Austria %\\This line break forced with \textbackslash\textbackslash
}%
\author{Georg Kresse}
 \altaffiliation[Also at ]{VASP Software GmbH.}%Lines break automatically or can be forced with \\
\affiliation{ 
Computational Materials Physics, Faculty of Physics, University of Vienna, 1090 Vienna, Austria %\\This line break forced with \textbackslash\textbackslash
}%

\date{\today}% It is always \today, today,
             %  but any date may be explicitly specified

\begin{abstract}
We present a method combining first-principles calculations and machine learning to predict the redox potentials of half-cell reactions on the absolute scale. By applying machine learning force fields for thermodynamic integration from the oxidized to the reduced state, we achieve efficient statistical sampling over a broad phase space. Furthermore, through thermodynamic integration from machine learning force fields to potentials of semi-local functionals, and from semi-local functionals to hybrid functionals using $\Delta$-machine learning, we refine the free energy with high precision step-by-step. Utilizing a hybrid functional that includes 25\% exact exchange (PBE0), this method predicts the redox potentials of the three redox couples, $\mathrm{Fe}^{3+}$/$\mathrm{Fe}^{2+}$, $\mathrm{Cu}^{2+}$/$\mathrm{Cu}^{+}$, and $\mathrm{Ag}^{2+}$/$\mathrm{Ag}^{+}$, to be 0.92, 0.26, and 1.99 V, respectively. These predictions are in good agreement with the best experimental estimates (0.77, 0.15, 1.98 V). This work demonstrates that machine-learned surrogate models provide a flexible framework for refining the accuracy of free energy from coarse approximation methods to precise electronic structure calculations, while also facilitating sufficient statistical sampling.
\end{abstract}

\maketitle

\section{\label{section1} Introduction}

Green energy and a circular economy are one of the key paradigms that our human society needs to realize in the next few decades. This implies that we need to give up on the combustion of fossil fuels. A key element to achieve this paradigm shift is the use of electrochemistry, be it for batteries and fuel cells, to convert electrical energy to hydrogen or other valuable chemicals, or to convert hydrogen back to energy without direct combustion in air.

The redox potential of electron transfer (ET), $\mathrm{Ox} + n\mathrm{e}^{-} \rightarrow \mathrm{Red}$ in liquids, is an essential property for a variety of electrochemical energy conversion devices, such as batteries, fuel cells, and electrochemical fuel synthesis. It determines the alignment of redox levels relative to the Fermi level of a metal, or valence band maximum (VBM) and conduction band minimum (CBM) of semiconductor and insulator electrodes. It also determines the stability windows of ions and molecules in solutions, that is the range of voltages within which a specific ion or molecule can undergo electrochemical reactions. This information is vital to design redox species and solvent molecules, such as redox couples for redox-flow batteries~\cite{WeberJApplElectrochm_2011}, solvents and additives for Li-ion batteries~\cite{Ong_ChemMater_2011, Xu_ChemRev_2014, Haregewoin_EnergyEnvironSci_2016}, radical scavengers for fuel cells~\cite{Zaton_SusEnergyFuel_2017} and electrocatalysts for fuel synthesis~\cite{Pinaud_EnergyEnvSci_2013, Morikawa_ACR_2022}. 

 \begin{table}[t]
\begin{center}
\caption{Redox potentials $U_{\mathrm{redox}}$ of three transition metal cations calculated by RPBE+D3, PBE0 (0.25), PBE0 (0.50), PBE0+D3 (0.25) and PBE0+D3 (0.50) using MLFF and $\Delta$-ML. Here, values in the parenthesis are the fraction of the exact exchange. The results for 64 water molecular systems are tabulated. The absolute potential of SHE is set to 4.44 V~\cite{Trasatti_PAC_1986}. The root means square errors (RMSE) compared to the experimental redox potentials are also shown. The results by the HSE06 functional reported by Sprik and co-workers~\cite{Liu_JPCB_2015} are also listed.} 
\label{table1}
\centering 
\begin{tabular}{p{25mm} p{13mm} p{13mm} p{13mm} p{13mm}}%
%\begin{tabular}{lcccc}	
\hline\hline	
XC functional & Fe  & Cu  & Ag  & RMSE \\
\hline
RPBE+D3                           &0.80                &0.66                                       &1.88                                        &0.29 \\    
PBE0 (0.25)                        &\textbf{0.92}     &\textbf{0.26}                           &\textbf{1.99}                             &0.11  \\
PBE0 (0.50)                        &0.79\                &$-$0.34                                  &2.12                                        &0.30  \\
PBE0+D3 (0.25)                   &\textbf{0.94}     &\textbf{0.24}                            &\textbf{2.02}                             &0.11  \\
PBE0+D3 (0.50)                   &0.83                &$-$0.38                                 &2.13                                          &0.32 \\
\hline
HSE06\textsuperscript{[a]}   &                      &$-$0.20      &1.72                &0.31 \\           
\hline                              
Exp.\textsuperscript{[b]}   &0.77          &0.15                                     &1.98                                          & \\
\hline
\end{tabular}
\end{center}
\footnotesize{\textsf{[a] From Ref.~\cite{Liu_JPCB_2015}. The calculations were done by the CP2K~\cite{Khne_JCP_2020} code with using the Goedecker-Teter-Hutter dual Gaussian norm-conserving pseudopotential~\cite{Goedecker_PRB_1996} with the electronic configuration of 3s$^{2}$3p$^{6}$3d$^{10}$4s$^{1}$ for Cu and 4s$^{2}$4p$^{6}$4d$^{10}$5s$^{1}$ for Ag and localized Gaussian basis sets for orbital.}}
\footnotesize{\textsf{[b] From Ref.~\cite{bard_book_1985}.}}
\end{table}

Unfortunately, to date, accurate first-principles (FP) predictions of this crucial property remain challenging, with typical prediction errors around 0.5 V. Sprik and co-workers developed a thermodynamic integration (TI) method utilizing the computational standard hydrogen electrode (CSHE)~\cite{Costanzo_JCP_2011, Le_PRL_2017} and applied this method to several redox reactions in aqueous solutions~\cite{Adriaanse_JCPL_2012, Liu_JPCB_2015}. They discovered that the use of a semi-local functional leads to errors exceeding 0.5 V. This discrepancy arises because the functional inaccurately yields the shallow valence band edge and the deep conduction band edge, resulting in incorrect hybridization with the redox levels. Similar magnitudes of errors have also been observed in other FP calculations that employ semi-local approximations~\cite{Caro_JCTC_2017, Bouzid_JCTC_2017}. As a result, Sprik and co-workers opted for a hybrid functional. Nonetheless, they observed a significant spread of values for two metal ion couples, with the $\mathrm{Cu}^{2+}$/$\mathrm{Cu}^{+}$ couple ranging from $-1.13$ to $-0.20$ V (experimental value 0.16 V) and the $\mathrm{Ag}^{2+}$/$\mathrm{Ag}^{+}$ couple ranging from 0.90 to 1.72 V (experimental value 1.98 V)~\cite{Liu_JPCB_2015}. These variations were attributed to differences in the pseudopotential and the computational code base (CMPD versus CP2K). While the "best" values obtained using the hybrid functional and highly accurate pseudopotentials are relatively close to experimental values ($-0.20$ V for Cu, and $1.72$ V for Ag), the agreement is still far from being quantitative. Due to the high computational cost of hybrid functionals, most calculations have been performed using approximated methods, such as continuum solvation models~\cite{Baik_JPCA_2002, Jaque_JPCC_2007, Jinnouchi_JPCC_2008, Neugebauer_JPCA_2020} and QM/MM models~\cite{Vaissier_JCTC_2016, Nicholson_JPCA_2021}. Although these models can reproduce the experimental redox potentials of ions and molecules with convenient accuracy, the computational results heavily rely on many approximations, making it unclear which predictions are strictly correct. Here, we briefly note that these FP and approximated methods have been extended to electrochemistry at liquid-solid interfaces~\cite{Taylor_PRB_2006, Skulason_PCCP_2007, Jinnouchi_PRB_2008, Letchworth-Weaver_PRB_2012, Kiran_JCP_2014, Otani_PRB_2017, Hormann_npj_2019, Gross_COE_2019, Islam_JCP_2023}. Nowadays, these methods have become indispensable for elucidating electrochemical interfacial phenomena and designing advanced materials~\cite{Norskov_JPCB_2004, Norskov_JES_2005, Norskov_NC_2009, Jinnouchi_PCCP_2011, Kulkarni_CR_2018, Nitopi_CR_2019}. However, even in the calculation of redox reactions at interfaces, approximations are made in most of applications, such as representing the motion of atomic nuclei with simple statistical models like the harmonic oscillator model~\cite{Norskov_JPCB_2004, Norskov_JES_2005, Taylor_PRB_2006, Skulason_PCCP_2007, Jinnouchi_PCCP_2011, Man_CCC_2011, Kortlever_JPCL_2015}, modeling solvents by molecular mechanics~\cite{Otani_PRB_2017}, or modeling by continuum mediums~\cite{Jinnouchi_PRB_2008, Letchworth-Weaver_PRB_2012, Kiran_JCP_2014, Hormann_npj_2019, Islam_JCP_2023}. A rigorous FP method that eliminates these approximations is also desired in the field of interfacial electrochemistry.

\begin{figure*}[t]
\begin{center}
\includegraphics[width=17.4cm]{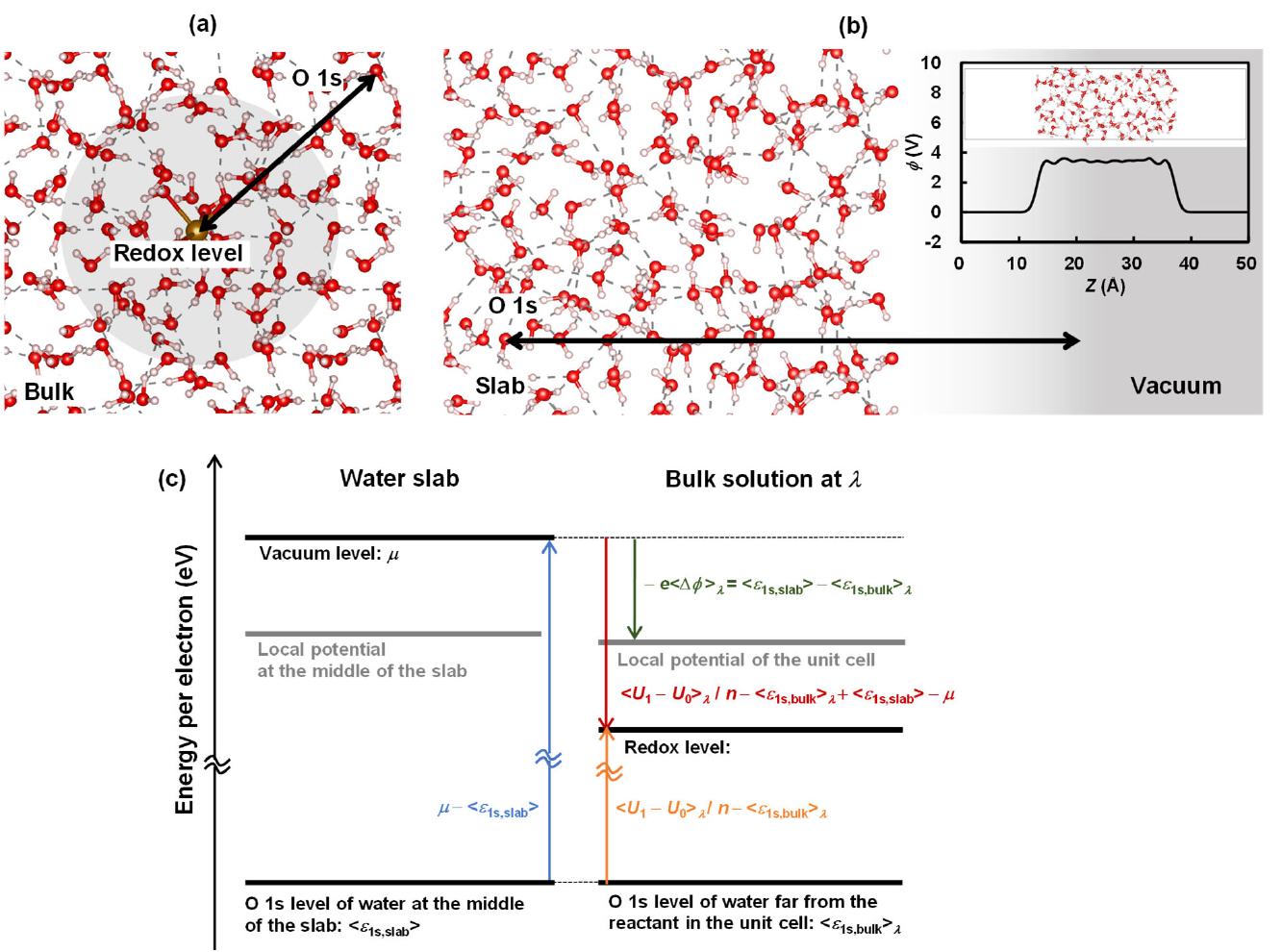}
\caption{Aligning energy levels based on the O 1s level of water molecules: (a) aligning the redox level based on the O 1s level of water molecules far from the redox species in the bulk solution model under a periodic boundary condition, (b) aligning the O 1s level of water molecules at the middle of the slab based on the local potential at the middle of the vacuum layer in the slab model, and (c) schematic of the alignment. The figure inset in (b) shows the snapshot of the water slab and computed local potential profile across the water slab. The graphics showing bulk and interfacial models are made by VESTA~\cite{Momma_JAC_2011}.
}
\label{fig1}
\end{center}
\end{figure*}

The main goals of the present work are three-fold: First, we want to accurately calculate the redox potential of metal ions in water for three prototypical cases: Ag, Cu, and Fe. Ag$^{2+}$ ions are among the most aggressive oxidants with a large redox potential, whereas the redox potential of Cu$^{2+}$ ions is fairly shallow, and the $\mathrm{Fe}^{3+}$/$\mathrm{Fe}^{2+}$ reaction lies in between. The first two redox reactions involve large changes in the ion  water coordination, which makes the calculation challenging, whereas the redox reaction of Fe is a so called simple outer sphere ET reaction and has been the subject of numerous experimental and theoretical studies~\cite{Marcus_RevModPhys_1993}. The Fe ions are conceived to be particularly challenging for density functional theory. Second, we want to establish a computationally feasible pathway that yields statistically accurate results. Last but not least, we want to systematically explore different density functionals to set a guideline for future studies.

\section{\label{section2}Overview of theory and modelling}

We begin with an overview of the used theory and modeling. Further details can be found in the Methods section. The reactions evaluated in this study are electron transfer reactions in water: $\mathrm{Fe}^{3+}+\mathrm{e}^{-} \leftrightarrow \mathrm{Fe}^{2+}$, $\mathrm{Cu}^{2+}+\mathrm{e}^{-} \leftrightarrow \mathrm{Cu}^{+}$, and $\mathrm{Ag}^{2+}+\mathrm{e}^{-} \leftrightarrow \mathrm{Ag}^{+}$. Here, we assume that other side reactions do not occur, and only the valency of redox species changes due to the reaction similarly to the previous study~\cite{Liu_JPCB_2015}.

The redox potential $U_{\mathrm{redox}}$ is determined by the Gibbs free energy difference $\Delta A$ between the reduced and oxidized states as
\begin{align}
U_\mathrm{redox} &= -\frac{\Delta A}{en}, \label{eq_1}
\end{align}
where $e$ is the elementary charge and $n$ is the number of electrons involved in the reaction. The free energy difference $\Delta A$ can be exactly determined by thermodynamic integration (TI)\cite{Zwanzig_JCP_1954, Kirkwood_JCP_1935}:
\begin{align}
\Delta A &= \int_{0}^{1} \left< \frac{\partial H}{\partial \lambda} \right>_{\lambda} d\lambda. \label{eq_2}
\end{align}
Here, $\langle X \rangle_{\lambda}$ denotes the expectation value of $X$ for an ensemble created by the Hamiltonian at coupling $\lambda$. The ensemble required for computing the Gibbs free energy difference is the isothermal-isobaric ensemble. However, in practice, due to the minor impact of volumetric change caused by the reaction, we utilized the canonical ensemble with the empirical density of liquid water, similar to previous publications~\cite{Costanzo_JCP_2011, Liu_JPCB_2015}. The integral seamlessly connects the oxidized state ($\lambda=0$) to the reduced state ($\lambda=1$) along a coupling path~\cite{Blumberger_JCP_2006, Dorner_PRL_2018}. The potential energy surface upon which atoms move is described by the grand potential $\Omega$ of the system opened for electrons~\cite{Mermin_PR_1965}. Consequently, the Hamiltonian of the system is described as follows:
\begin{align}
H &= \sum\limits_{i=1}^{N_{\mathrm{a}}}  \frac{ \left| \mathbf{p}_{i} \right|^{2}}{2 m_{i}} + \Omega, \label{eq_3} \\
\Omega &= U - \mu N , \label{eq_4} 
\end{align}
where $N_\mathrm{a}$ is the number of atoms, $m_{i}$ and $\mathbf{p}_{i}$ are the mass and momentum vector of $i$-th atom, and $\mu$ and $N$ are the chemical potential and the number of electrons. The chemical potential $\mu$ is fixed at the reservoir level, whereas $N$ varies by $n$ along the coupling path. $U$ represents the potential energy surface at $\lambda$, equating to the sum of the Helmholtz free energy of the electronic subsystem and the electrostatic interactions among nuclei. Following previous studies~\cite{Blumberger_JCP_2006, Dorner_PRL_2018}, $U$ can be described as
\begin{align}
U &= \lambda U_{1} + \left( 1- \lambda \right) U_{0}, \label{eq_5} 
\end{align}
where $U_{0}$ and $U_{1}$ are the potential energies of the oxidized and reduced states, respectively. Hence, the free energy difference $\Delta A$ is written as
\begin{align}
\Delta A &= \int_{0}^{1} \left< U_{1} - U_{0} \right>_{\lambda} d\lambda - \mu n. \label{eq_6}
\end{align}
If the structural changes are significant from the oxidized to the reduced species --- recall  that this is the case for Ag and Cu --- many integration steps are required to accurately determine the energy difference. The application of this approach entails two difficulties. (i) Clearly, it implies huge computational cost if applied directly to hybrid functionals; if 100.000 timesteps using a complete plane wave basis set are required to obtain good statistical accuracy, several 10 mio core hours are necessary. (ii) Second, during the reaction one electron needs to be transferred from the reservoir, characterizing the chemical potential of the electrons. The vacuum level is the best suited reference chemical potential that allows one to align the redox levels and band edges of the electrode in the absolute potential scale. However, in FP calculations of bulk systems under periodic boundary conditions, the vacuum level is a quantity that cannot be directly accessed during simulations.

{\em Chemical potential of electrons:} We will address the second point (ii) first. Jiao and co-workers~\cite{Jiao_JCTC_2011} suggested to use the average electrostatic potential as suitable reference point, and Leung~\cite{Leung_JPCL_2010} calculated the position of the average electrostatic potential with respect to the vacuum level in a second independent calculation involving a water slab. We refine this approach in a conceptually easy to understand way that simultaneously reduces finite size errors. As a reference, instead of using the vacuum level, we employ the O 1s level of water, which is fixed relative to the vacuum level and can be conveniently calculated with the FP code used in this study. Our approach is schematically illustrated in Fig.\ref{fig1}. In FP calculations of a solution system under a periodic boundary condition, the energy contribution $\left< U_{1}-U_{0} \right>_{\lambda}$ in Eq.(\ref{eq_6}) is equal to the negative electron affinity of the oxidized species scaled to the average local potential of the system. The same calculation can also determine the O 1s level $\left< \epsilon_\mathrm{1s,bulk} \right>_{\lambda}$ of water, sufficiently far from the redox species and unaffected by the reactant, scaled to the average local potential. Therefore, measuring the redox level using the O 1s level as a reference results in $\left< U_{1}-U_{0} \right>_{\lambda}/n - \left< \epsilon_\mathrm{1s,bulk} \right>$, as highlighted in orange letters in Fig.\ref{fig1} (c). In practice, $\left< \epsilon_\mathrm{1s,bulk} \right>_{\lambda}$ may slightly vary along the coupling path due to finite size effects [refer to Table S4 in the Supporting Information (SI)]. By aligning the potentials between the 'defect' and the 'host' within the same supercell in this manner, the finite size effects can be mitigated~\cite{Lany_PRB_2008, Freysoldt_RMP_2014}. The vacuum level referenced to the O 1s level can be calculated using a slab model. As depicted in Fig.\ref{fig1} (b), when referencing the O 1s level of water molecules located in the middle layer of the water slab, the vacuum level can be expressed as $\mu - \left< \epsilon_\mathrm{1s,slab} \right>$, as indicated in blue letters in Fig.\ref{fig1} (c). The difference between the redox level and vacuum level scaled to the O 1s level results in the redox level scaled to the vacuum level, as shown in red letters in Fig.\ref{fig1} (c). Consequently, the free energy difference $\Delta A$ on an absolute scale is written as
\begin{align}
\Delta A &= \int_{0}^{1} \left< U_{1} - U_{0} \right>_{\lambda} d\lambda - n e \Delta \bar \phi,  \label{eq_7}  \\
e\Delta \bar \phi &= \int_{0}^{1} \left< \epsilon_\mathrm{1s,bulk} \right>_{\lambda} d\lambda - \left< \epsilon_\mathrm{1s,slab} \right>, \label{eq_8}
\end{align}
where the vacuum level $\mu$ is set to zero. As illustrated by the green letters in Fig.~\ref{fig1} (c), $\Delta \bar \phi$ accounts for the difference between the local potential at the vacuum in the slab model and the one in the bulk solution model.

Equation~(\ref{eq_7}) is similar to the one used for the computational standard hydrogen electrode (CSHE) method proposed in the previous studies~\cite{Costanzo_JCP_2011, Le_PRL_2017}. However, there are three differences in our approach compared to the previous method. Firstly, our approach computes the absolute vacuum reference instead of referencing to the standard hydrogen electrode (SHE), thereby enabling the evaluation of absolute potentials of half-cell reactions. Second, aligning using the O 1s level in the same supercell is expected to mitigate the finite size effects. Third, machine learned (ML) force fields (FFs) can create many statistically independent configurations for the water slab. We do this by on-the-fly learning an H$_2$O force field for the bulk and then for the surface, and performing finally extensive million step (total 1.5 ns) ML molecular dynamics for the surface. From this simulation we draw 3000 statistically independent snapshots. Only for these snapshots, FP calculations are performed to determine the average O 1s level with respect to the vacuum level. This substantially reduces the required computational time from 1 mio core hours for brute force runs using the semi-local functional to only 2200 core hours for the FP calculations on 3000 structures, including the ML simulations and training runs, while retaining statistical accuracy, as demonstrated by the local potential profile shown in the inset of Fig.~\ref{fig1}(b) [see details of the estimation of compute time in Section S2 in SI].

\begin{figure}[t]
\begin{center}
\includegraphics[width=8.6cm]{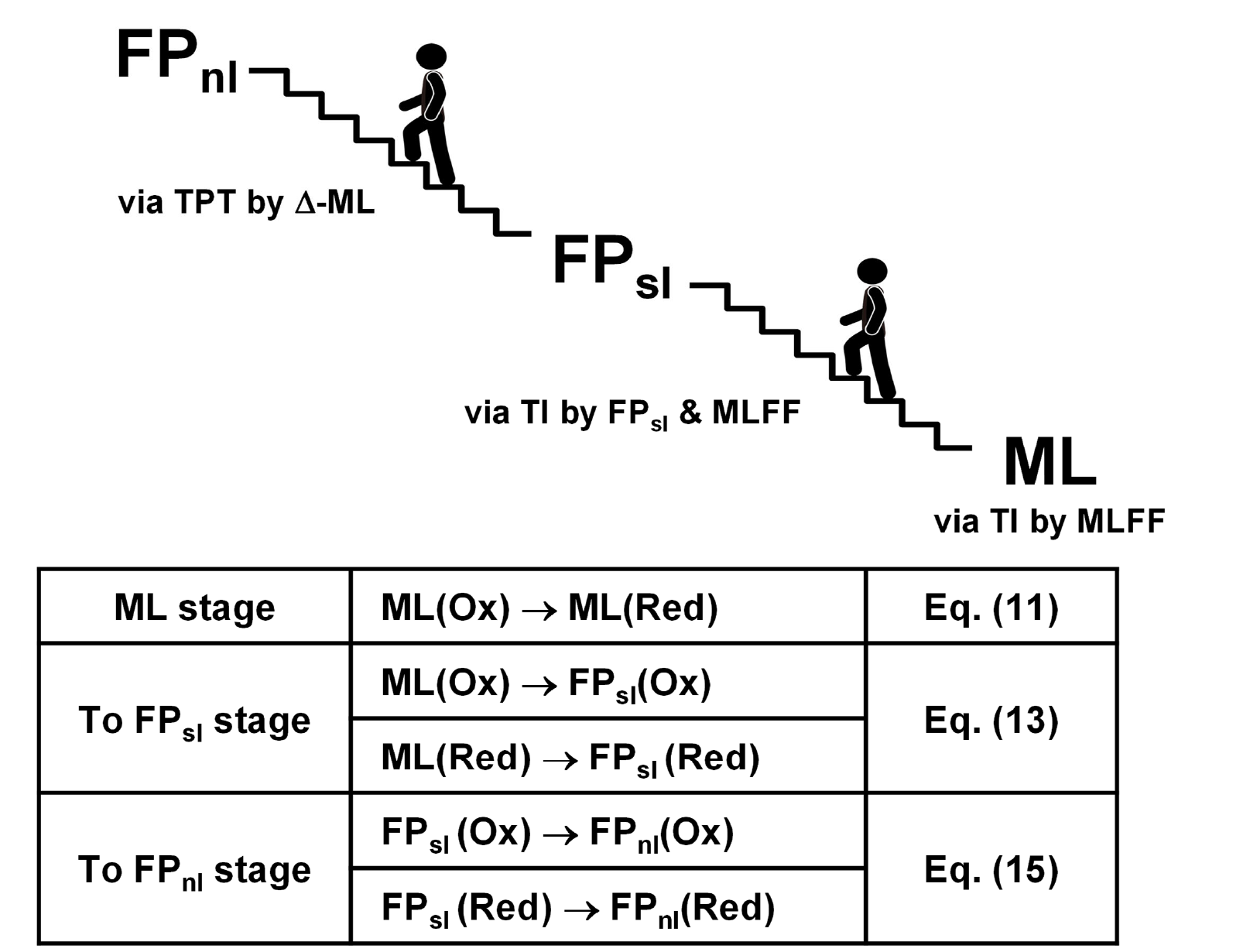}
\caption{Schematic of ML-aided TI and TPT to compute the free energy difference $\Delta A$. The notations ML, FP$_\mathrm{sl}$ and FP$_\mathrm{nl}$ mean the machine-learning force field, FP method with semi-local functional, FP method with non-local hybrid functional, respectively. See details of equations in the Methods section.
}
\label{fig2}
\end{center}
\end{figure}

{\em Thermodynamic integration:}
To address the problem to compute the free energy difference, i.e. point (i), we propose the ML-aided scheme as depicted in Fig.~\ref{fig2}. Here, we use the abbreviations $\mathrm{FP_{nl}}$(Ox/Red), $\mathrm{FP_{sl}}$(Ox/Red), and ML(Ox/Red) to denote calculations using a {\em non-local} hybrid functional, a {\em semi-local} functional and {\em machine-learned} force field for the oxidized and reduced cases, respectively. Naively, one could just perform the required TI using ML surrogate models. As we will show in Results and Discussion, this yields only acceptable accuracy. Errors in $\Delta A$ resulting from inaccuracies in the trajectory and the energy predictions by the ML potential can be corrected by performing TI from the ML potential to the FP potential for both the oxidized and reduced states. We will adopt this strategy for the $\mathrm{FP_{sl}}$ method. So this involves two calculations: TI from the oxidized to the reduced species using ML surrogate models via Eq.~(\ref{eq_11}) in Methods section, $\mathrm{ML}$ (Ox) $\rightarrow$ $\mathrm{ML}$ (Red), and then for each oxidation state, TI from MLFF to the FP$_\mathrm{sl}$ Hamiltonian via Eq.~(\ref{eq_13}), $\mathrm{ML}$(Ox)$\rightarrow$$\mathrm{FP_{sl}}$ (Ox) and $\mathrm{ML}$(Red)$\rightarrow$$\mathrm{FP_{sl}}$ (Red). This two-step integration has three advantages as summarized below:
\begin{itemize}
\item The integration $\mathrm{ML}$ (Ox) $\rightarrow$ $\mathrm{ML}$ (Red) using the MLFF takes into account most of the non-linear components of the integrand in the TI (see Figs. S8 and S9 in SI). Excellent statistical accuracy can be reached for this step.
\item The MLFFs also provide well-equilibrated  initial structures required for other calculational steps.
\item The integrands in $\mathrm{ML}$(Ox)$\rightarrow$$\mathrm{FP_{sl}}$(Ox) and $\mathrm{ML}$(Red)$\rightarrow$$\mathrm{FP_{sl}}$ (Red) are small and almost linear in the coupling parameter (see Fig. S10) owing to the accurate reproduction of the $\mathrm{FP_{sl}}$ structures by the MLFF (see Results and Discussion). Hence, these demanding integrals (evaluation of $\mathrm{FP_{sl}}$ calculation in every MD step) converge using a few tens of pico second MD simulations.
\end{itemize}

There is one final obstacle though: performing TI to a potential calculated by a hybrid functional that includes non-local exchange (FP$_\mathrm{nl}$) is still exceedingly demanding and challenging. So in this specific case, as depicted in Fig.~\ref{fig2}, we have decided to apply the $\Delta$-machine learning ($\Delta$-ML)~\cite{Balabin_JCP_2009, Ramakrishnan_JCTC_2015, Bartok_PRB_2013, Chmiela_NatCommun_2018, Sauceda_JCP_2019, Liu_2022_PRB, Carla_PRM_2023, Liu_PRL_2023} which learns the {\em difference} $\Delta U$ between the FP$_\mathrm{sl}$ potential and the FP$_\mathrm{nl}$ potential. Due to the very smooth energy difference between the FP$_\mathrm{sl}$ functional and the FP$_\mathrm{nl}$ functional, it is possible to learn an extremely accurate ML representation of $\Delta U$ with just a few tens of FP$_\mathrm{nl}$ calculations, with errors significantly smaller by an order of magnitude or more compared to those associated with MLFF models (see details in Figs. S2 to S4 and Fig. S17). In the current implementation, the TI integration has been replaced with thermodynamic perturbation theory (TPT),
\begin{align}
\Delta A &= A_1 - A_0 = -\frac{1}{\beta}\mathrm{ln} \left< e^{-\beta \Delta U} \right> _0  = -\frac{1}{\beta}\mathrm{ln} \left< e^{\beta \Delta U} \right> _1, \label{eq_10}
\end{align}
where $\beta$ is the inverse temperature, and the symbol $\Delta U$ denotes the potential energy difference between the two end points. Although Eq. (\ref{eq_10}) is in principle exact, the potential energy difference might need to be evaluated for thousands or even many ten thousand configurations if the ensembles generated by the two potentials are too distinct. This implies the significantly expensive FP$_\mathrm{nl}$ calculations. The $\Delta$-ML scheme allows for the circumvention of this issue, enabling the reduction of the required FP$_\mathrm{nl}$ calculations to merely tens. Thanks to the remarkable accuracy of the $\Delta$-ML models, it is possible to obtain exceedingly accurate free energy differences between different FP methods without further correction (see Fig. S12 in SI). This is one of the key advances of the present work. The computational cost is ultimately only limited by generating sufficient configurations using the FP$_\mathrm{sl}$. Thus, the required compute time for direct TI using the FP$_\mathrm{nl}$ method is reduced from 20 mio core hours to 16800 core hours for the FPMD simulations that generate configurations using the FP$_\mathrm{sl}$ method (see details of the estimation in Section S2 in SI).

\section{\label{section3}Results and discussion}

\begin{figure}[t]
\begin{center}
\includegraphics[width=8.6cm]{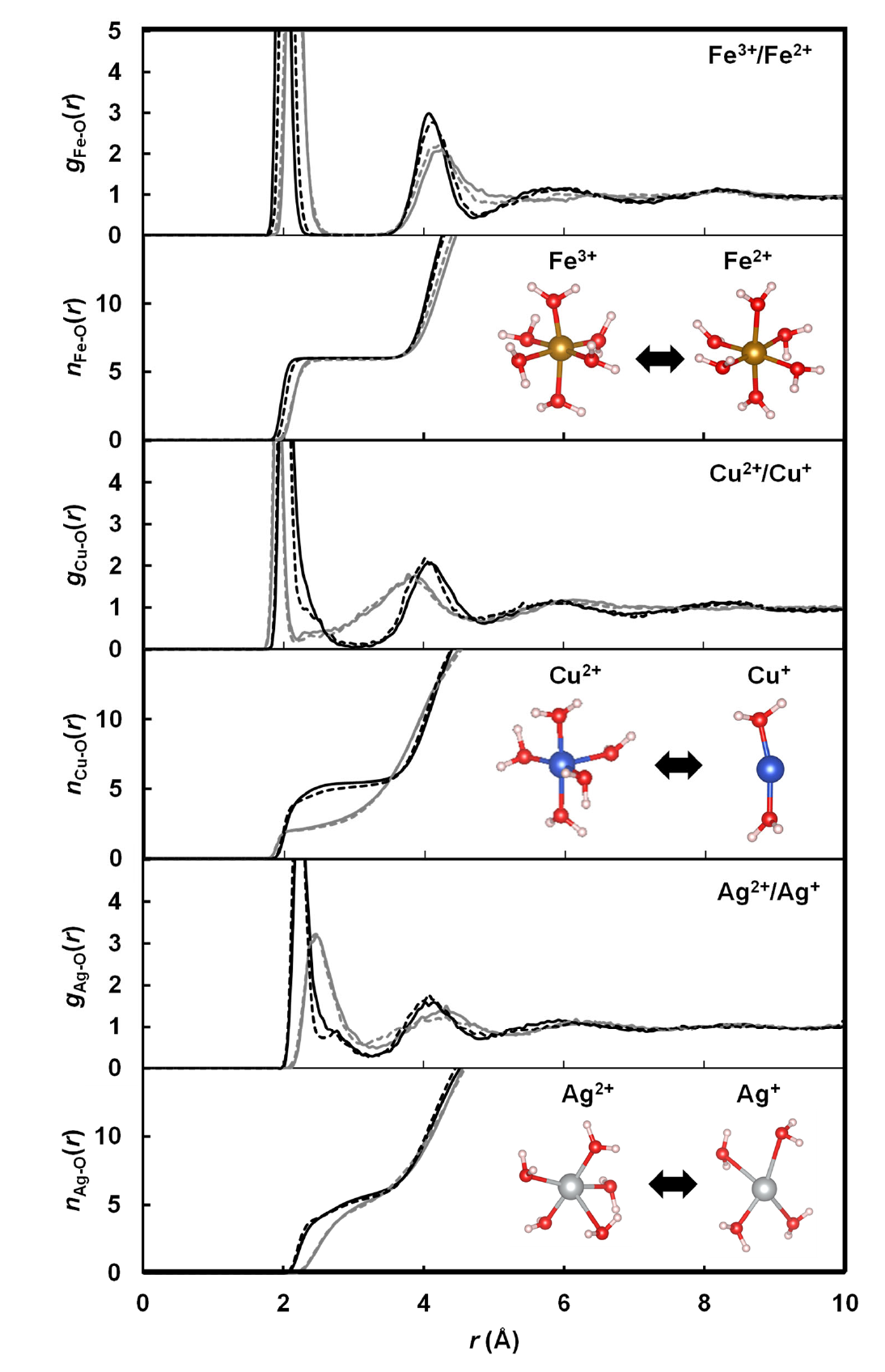}
\caption{Metal-oxygen radial distribution functions ($g_{\mathrm{X-O}}$) and running integration numbers ($n_{\mathrm{X-O}}$) provided by 100 ps MLFF-MD and 10 ps FPMD simulations. Gray and black lines are for the reduced and oxidized states, respectively. Solid and dashed lines are results obtained by the MLFF and FP$_{\mathrm{sl}}$, respectively. Graphics in the insets show first solvation structures at the reduced and oxidized states.}
\label{fig3}
\end{center}
\end{figure}

We now detail our results, and will show that the adopted procedure yields statistically highly accurate results. The calculations were performed using VASP~\cite{Kresse_PRB_1996, Kresse_CMS_1996} and the projector augmented wave (PAW) method~\cite{Blochl_PRB_1994, Kresse_PRB_1999}. For the ML force fields (MLFFs) the implementation detailed in previous publications is used~\cite{Jinnouchi_PRB_2019, Jinnouchi_JPCL_2020, Jinnouchi_JPCL_2023}. Similar to the pioneering ML approaches~\cite{Behler_PRL_2007, Bartok_PRL_2010, Bartok_PRB_2013}, the potential energy in our MLFF method is approximated as a summation of local energies [see Eq.~(\ref{eq_20})].  The local energy is approximated as a weighted sum of kernel basis functions [see Eq.~(\ref{eq_21})]. A Bayesian formulation allows to efficiently predict energies, forces and stress tensor components as well as their uncertainties. The predicted uncertainty enables the reliable sampling of the reference structures on-the-fly during the FPMD simulation. Details of the equations, parameters and training conditions are summarized in the Methods section and Section S1 in SI. As in the previous studies~\cite{Jinnouchi_PRL_2019, Jinnouchi_PRB_2019, Jinnouchi_PRB_2020, Jinnouchi_JCP_2020}, the MLFFs trained on a semi-local functional with dispersion corrections achieve root mean square errors (RMSEs) of 1-5 meV atom$^\mathrm{-1}$ and 0.04-0.11 eV \AA$^\mathrm{-1}$ for energies and forces (see error distributions in Figs. S1 to S4 in SI). 
The three ET reactions are examined in water by using a semi-local functional~\cite{Hammer_PRB_1999} with a dispersion correction~\cite{Grimme_JCP_2010, Grimme_JCC_2011} (RPBE+D3) and hybrid functionals~\cite{Perdew_JCP_1996, Adamo_JCP_1999} with and without a dispersion correction (PBE0 and PBE0+D3). Systematic comparisons of different functionals help us to study  the effects of the exact exchange as well as dispersion corrections. As shown in Table~\ref{table1} [see lines of PBE0 (0.25) and PBE0+D3 (0.25)] good agreement with experiment is achieved using the hybrid functional with 1/4 exact exchange, regardless of whether dispersion corrections are used or not.

\begin{figure*}
\begin{center}
\includegraphics[width=15.8cm]{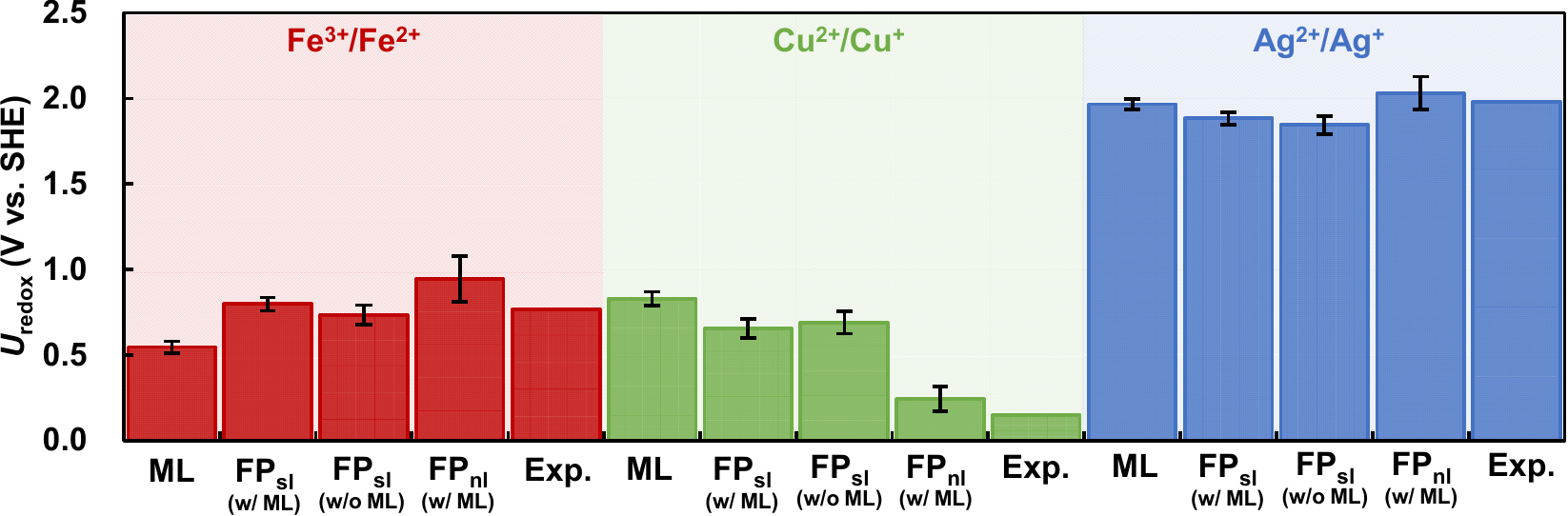}
\caption{Computed and experimental redox potentials. ML means the results obtained by the MLFF trained on $\mathrm{FP_{sl}}$ (RPBE+D3) without any correction. The small letters `w/ ML' under $\mathrm{FP_{sl}}$ mean that the ML result was corrected by the scheme shown in Fig.~\ref{fig2}. The letters `w/o ML' mean the result calculated by $\mathrm{FP_{sl}}$ using Eqs.~(\ref{eq_18}) and (\ref{eq_19}) without MLFF. Here, $\mathrm{FP_{nl}}$ means PBE0+D3 (0.25) result obtained by the scheme shown in Fig.~\ref{fig2}.  Experimental values are taken from Ref.~\cite{bard_book_1985}.}
\label{fig4}
\end{center}
\end{figure*} 

{\em Water surface calculations:}
For RPBE+D3, the present MLFF  provides a surface tension of 79$\pm$5 mN m$^{-1}$ for the 128 molecular system and 84$\pm$5 mN m$^{-1}$ for the 1024 molecular system at 298 K. Here, the surface tension was computed as~\cite{Wohlfahrt_JCP_2020}
\begin{align}
\gamma &= \frac{L_{z}}{2} \left( p_{zz} - \frac{p_{xx} + p_{yy}}{2} \right), \label{eq_11}
\end{align}
where $x$ and $y$ define the directions parallel to the macroscopic interface, $z$ defines the direction normal to the interface, $L_{z}$ is the length of the unit cell in the $z$-direction, and $p_{ij}$ are the pressure tensor. The results are slightly larger than the value of 68$\pm$2 mN m$^{-1}$ calculated by a neural network potential~\cite{Wohlfahrt_JCP_2020} and experimental value of 72 mN m$^{-1}$~\cite{Vargaftik_JCP_1983} while it is within the range (50-90 mN m$^{-1}$) of previous MD results by FP~\cite{Ohto_JCTC_2019} and classical force fields~\cite{Taylor_JCP_1996, Vega_JCP_2007}. Distributions of interfacial water dipole moments for both, 128 and 1024 molecular systems, are shown in Fig. S5. They consistently indicate that the orientation of interfacial water molecules is bimodal as reported in previous MD studies employing the classical SPC/E force field~\cite{Taylor_JCP_1996}. The distributions are also consistent with the results of sum frequency generation (SFG) analyses~\cite{Du_PRL_1993}. 

{\em Metal water coordination:}
Figure~\ref{fig3} shows metal-oxygen radial distribution functions (RDFs) and running integration numbers (RINs) at the reduced and oxidized states calculated by the MLFF and FP$_{\mathrm{sl}}$ methods. The MLFFs well reproduce RDFs and RINs of the FP$_{\mathrm{sl}}$ method. Both methods show that the coordination number of Fe ions is 6 independent of the charge state. In contrast, the value for Cu changes from 5-6 in the oxidized state (Cu$^{2+}$) to 2-3 in the reduced state (Cu$^{+}$). The coordination number of Ag also changes from 5-6 in the oxidized state (Ag$^{2+}$) to 4-5 in the reduced state (Ag$^{+}$). These hydration structures agree with the ones reported in previous MD studies using FPMD methods~\cite{Remsungen_CPL_2004, Blumberger_JACS_2004, Bogatko_JPCA_2010} and empirical force fields~\cite{Remsungen_CPL_2004}. Although there are slight deviations in the Fe-O distance and shoulders for Cu-O and Ag-O in the RDFs likely related to the short FPMD simulation time and errors in the MLFFs, overall, our MLFFs reproduce the first-principles energies and structures of the hydrated metal cations with good accuracy.

{\em Redox potentials:}
After verifying the size effect on the redox potentials obtained on the $\mathrm{FP_{sl}}$ level with using unit cells containing 32, 64 and 96 water molecules (see $U_\mathrm{redox}$ in Fig. S13), calculations were conducted on the bulk solutions containing 64 water molecules in the unit cell. The computed redox potentials are compared with the experimental ones in Fig.~\ref{fig4}. All relevant data ($\left< U_{1}-U_{0} \right>_{\lambda}$ and $\Delta \bar \phi$), as well as results of other functionals with error bars, are summarized in Sections S3, S4 and S5. The MLFFs trained on $\mathrm{FP_{sl}}$ (RPBE+D3) (see ML in Fig.~\ref{fig4}) lead to non-negligible deviations of 30-250 mV from the values of a full $\mathrm{FP_{sl}}$ calculations without any MLFF (see $\mathrm{FP_{sl}}$ w/o ML) depending on the training data size (see Section S9 in SI). The deviations can be corrected by two TI integrations [$\mathrm{ML}$(Ox) $\rightarrow$ $\mathrm{FP_{sl}}$(Ox) and $\mathrm{ML}$(Red) $\rightarrow$ $\mathrm{FP_{sl}}$(Red) in Fig.~\ref{fig2}] as shown by $\mathrm{FP_{sl}}$ w/ ML. However, the semi-local functional result in fairly large and non-systematic errors. For Ag the redox potential is underestimated, whereas for Cu it is significantly overestimated compared to experiment.

The errors can be significantly decreased to 0.11 V on average  using hybrid functionals with one quarter exact exchange. As tabulated in Table~\ref{table1}, $U_\mathrm{redox}$ for the Cu$^\mathrm{2+}$/Cu$^\mathrm{+}$ couple decreases with increasing fraction of the exact exchange, whereas the redox potential for the Ag$^\mathrm{2+}$/Ag$^\mathrm{+}$ couple increases with increasing the fraction.  For Fe$^\mathrm{3+}$/Fe$^\mathrm{2+}$, the trend is not so obvious (first increase then slight decrease). Overall the present trends agree with the results obtained using semi-local and hybrid functionals as reported by Liu and co-workers~\cite{Liu_JPCB_2015}. Finally, the effects of Grimme's dispersion correction is small for all redox couples. This implies that changes of the electronic properties (such as water valence band maximum and conduction band minimum) are most relevant, whereas all the functionals give a similar and good account of the solvation structure. It remains unclear, however, why one quarter of exact exchange results in balanced accuracy. The functional form of PBE0 was rationalized by the adiabatic connection from the uncorrelated exact exchange to the fully interacting energy, which is approximated by the PBE functional~\cite{Becke_JCP_1993, Perdew_JCP_1996}. Nonetheless, the ratio of exact exchange continues to be a parameter. One quarter of exact exchange is known to achieve balanced accuracy for the geometries, thermochemistry, and spectroscopic properties of molecules. However, as reported in previous studies~\cite{Pham_PRB_2014}, this functional underestimates the band gap of liquid water, even though it provides a more accurate prediction than the HSE06 functional. The mechanism behind this remains an open question.

Another key observation in this study lies in the relationship between the error of the ML surrogate model and the error in the redox potential. Our MLFF models achieve an RMSE of a few meV atom$^{-1}$ for energy and tens of meV \AA$^{-1}$ for force. These accuracies can be considered standard level compared to ML models generated in past research~\cite{Behler_PRL_2007, Bartok_PRL_2010, Bartok_PRB_2013, Jinnouchi_PRB_2019, Jinnouchi_JCP_2020, Shapeev_MMS_2016, Drautz_PRB_2019,Lysogorskiy_npj_2021}, yet they yield non-negligible deviations in the redox potential from the FP method. In comparison, $\Delta$-ML models, which attained a RMSE substantially lower by more than an order of magnitude, markedly diminished the deviation in the redox potential to below 10 mV (refer to Fig. S12 in the SI). These results suggest that, in aiming for an accuracy of 10 mV in reproducing the redox potential of the FP method, an RMSE at least an order of magnitude smaller than that shown by standard MLFFs is required. Achieving this level of accuracy is highly challenging for MLFFs, even if they are trained on larger training datasets, as demonstrated in the previous study on liquid water~\cite{Hijes_JCP_2024}. While the accuracy of emerging MLFFs continues to improve~\cite{Batzner_NatCommn_2022, Ilyes_NEURIPS_2022}, there is always a risk that machine learning models may produce errors concerning the structure of extrapolation regions outside the training data. Even in a future where machine learning models have further advanced, our ML correction schemes will serve as a powerful method for quantifying errors and providing results from accurate FP calculations.

In summary, our approach enables efficient statistical sampling that is indispensable for accurate computations of the free energies of aqueous systems. The TI and TPT calculations allow to improve the accuracy from the ML model to the semi-local functional and from the semi-local functional to the  hybrid functional step-by-step.
Combining TPT and $\Delta$-machine learning are particularly promising, since this allows to obtain statistically highly accurate results even for expensive functionals in very little compute time. For instance, it is well conceivable that one could also use methods beyond density functional theory for the final step.
Our final results reproduce the redox potentials of the three transition metal cations with excellent accuracy using a standard hybrid functional. 
The integration pathways chosen here are generalizable to a wide variety of electron transfer reactions. We believe that the scheme will pave the way to first-principles electrochemistry predicting the key property of redox reactions in energy conversion devices.

\section{\label{section4}Methods}

\subsection{TI and TPT}

The TI and TPT shown in Fig.~2 in the main text is conducted by using the equations listed below.
\begin{itemize}
\item $\mathbf{ML(Ox) \rightarrow ML(Red)}$
\end{itemize}
\begin{align}
\Delta A^{\mathrm{ML}} &= \int_{0}^{1} \left<  \frac{ \partial H^{\mathrm{ML}}}{\partial \lambda}  \right>_{\lambda} d\lambda, \label{eq_11} \\
H^{\mathrm{ML}}&=\sum\limits_{i=1}^{N_{\mathrm{a}}} \frac{ \left| \mathbf{p}_{i} \right| ^{2}}{2 m_{i}} + \lambda U_{1}^{\mathrm{ML}} + \left( 1-\lambda \right) U_{0}^{\mathrm{ML}} - Ne\Delta \bar\phi. \label{eq_12}
\end{align}
\begin{itemize}
\item $\mathbf{ML(Ox) \rightarrow FP_{sl}(Ox)\; and\; ML(Red) \rightarrow FP_{sl}(Red)}$
\end{itemize}
\begin{align}
\Delta A_{\kappa}^{\mathrm{FP_{sl}-ML}} &= \int_{0}^{1} \left< \frac{\partial H_{\kappa}^{\mathrm{FP_{sl}-ML}}}{\partial \eta} \right>_{\eta} d\eta, \label{eq_13} \\
H_{\kappa}^{\mathrm{FP_{sl}-ML}} &= \sum\limits_{i=1}^{N_{\mathrm{a}}}  \frac{ \left| \mathbf{p}_{i} \right|^{2}}{2 m_{i}} + \eta U_{\kappa}^{\mathrm{FP_{sl}}} + \left( 1-\eta \right) U_{\kappa}^{\mathrm{ML}}. \label{eq_14}
\end{align}
\begin{itemize}
\item $\mathbf{FP_{sl}(Ox)  \rightarrow FP_{nl}(Ox) \;and\; FP_{sl}(Red)  \rightarrow FP_{nl}(Red) }$
\end{itemize}
\begin{align}
\Delta A_{\kappa}^{\mathrm{FP_{nl}-FP_{sl}}}  &\simeq \left< \Delta U_{\kappa}^{\mathrm{\Delta ML}} \right>_{\mathrm{FP_{sl}}} \nonumber \\
&- \frac{\beta}{2} \left< \left( \Delta U_{\kappa}^{\mathrm{\Delta ML}} - \left< \Delta U_{\kappa}^{\mathrm{\Delta ML}} \right>_{\mathrm{FP_{sl}}} \right)^{2} \right>_{\mathrm{FP_{sl}}}.  \label{eq_15}
\end{align}
The symbols $U_{\kappa}^\mathrm{FP_{nl}}$, $U_{\kappa}^\mathrm{FP_{sl}}$ and $U_{\kappa}^\mathrm{ML}$ are the potential energies for the oxidized ($\kappa=0$) and reduced ($\kappa=1$) states calculated by the non-local functional, semi-local functional and MLFF trained on the semi-local functional, respectively. The symbol $\Delta U_{\kappa}^{\mathrm{\Delta ML}}$ denotes the potential energy difference calculated by the $\Delta$-ML model trained on the potential energy difference $U_{\kappa}^\mathrm{FP_{nl}}-U_{\kappa}^\mathrm{FP_{sl}}$ between the non-local and semi-local functionals. In Eq.~(\ref{eq_15}), the second-order cumulant expansion is employed. The expansion is exact if the probability distribution of $\Delta U_{\kappa}^{\mathrm{\Delta ML}}$ is Gaussian (see derivation in Section S8 in SI). The condition is reasonably satisfiled as shown in Fig. S11. Preliminary TI and TPT simulations using the MLFFs also indicate that the TPT calculation reproduces TI results as shown in Section S6.

The free energy differences of the $\mathrm{FP_{sl}}$ and $\mathrm{FP_{nl}}$ methods are obtained as
\begin{align}
\Delta A^\mathrm{FP_{sl}} &= \Delta A^\mathrm{ML}  + \Delta A_{1}^{\mathrm{FP_{sl}-ML}} - \Delta A_{0}^{\mathrm{FP_{sl}-ML}}, \label{eq_16} \\
\Delta A^\mathrm{FP_{nl}} &= \Delta A^\mathrm{FP_{sl}} + \Delta A_{1}^{\mathrm{FP_{nl}-FP_{sl}}} -\Delta A_{0}^{\mathrm{FP_{nl}-FP_{sl}}}. \label{eq_17}
\end{align}
%where $\Delta A^\mathrm{ML}$ is the free energy difference computed by the $\mathrm{ML}$ method.

To validate the MLFF-aided computations of the free energy difference $\Delta A^\mathrm{FP_{sl}}$, the same property was also computed by the TI without using the $\mathrm{ML}$ method:
\begin{align}
\Delta A^{\mathrm{FP_{sl}}} &= \int_{0}^{1} \left<  \frac{ \partial H^{\mathrm{FP_{sl}}}}{\partial \lambda}  \right>_{\lambda} d\lambda, \label{eq_18} \\
H^{\mathrm{FP_{sl}}}&=\sum\limits_{i=1}^{N_{\mathrm{a}}} \frac{ \left| \mathbf{p}_{i} \right| ^{2}}{2 m_{i}} + \lambda U_{1}^{\mathrm{FP_{sl}}} + \left( 1-\lambda \right) U_{0}^{\mathrm{FP_{sl}}} - Ne\Delta \bar\phi. \label{eq_19}
\end{align}

The TI calculation in Eq.~(\ref{eq_2}) can be decomposed into the two terms on the right-hand side of Eq.~(\ref{eq_7}). The integrand in the first term nonlinearly varies along the coupling path (see Figs. S8 and S9 in SI) while the integrand in Eq.~(\ref{eq_8}), which is relevant to the second term in Eq.~(\ref{eq_7}), varies only slightly (see Table S4 in SI). To perform the integration of the first term in Eq.~(\ref{eq_7}), the Simpson's rule with equidistant five points was used following the previous FP study by Blumberger and co-workers~\cite{Blumberger_JCP_2006}. For the integration in Eq.~(\ref{eq_8}), the average of the O 1s levels in the fully reduced and oxidized states was used based on the trapezoidal rule. For each point, the ensemble average was taken over an 80-ps-NVT-ensemble MD simulation at 298 K after a 20 ps equilibration. Similar to the MLFF calculations, the Simpson's rule with equidistant five points was used for the TI calculation in Eq.~(\ref{eq_18}). For each grid, the ensemble average was taken over a 20-ps-MD simulation starting from the final structure of the TI calculation using the MLFF at the same grid point. Each initial structure of the MD simulations was prepared by annealing the system from 400 K to 298 K by a 100-ps-NVT-ensemble MD simulation using the MLFF after annealing the same system from 1000 K to 400 K by a 1-ns-NVT ensemble MD simulation using the polymer consistent force field (PCFF)~\cite{Sun_JCC_1994} implemented in a homemade MD program~\cite{Jinnouchi_ECA_2016}. Figures~S8 and S9 in SI show the integrands of Eqs.~(\ref{eq_11}) and~(\ref{eq_18}), respectively, as functions of the coupling parameter $\lambda$. In the same figures, probability distributions of $\Delta U^{\mathrm{ML}}=U_{1}^\mathrm{ML}-U_{0}^\mathrm{ML}$ and $\Delta U^{\mathrm{FP_{sl}}}=U_{1}^\mathrm{FP_{sl}}-U_{0}^\mathrm{FP_{sl}}$ at each $\lambda$ are also shown. For all redox couples, the variance of the distribution varies with changing $\lambda$, and thus, the integrand is non-linear with respect to $\lambda$ [see Eq.~(S14) in SI]. Hence, the second cumulant expansion Eq.~(S12) is not applicable to the whole integration from the oxidized state to the reduced state.

The TI calculations in Eq.~(\ref{eq_13}) were conducted using the trapezoidal rule with three equidistant points. At each point, a 10-ps-NVT-ensemble MD simulation at 298 K was performed. The integrands shown in Fig.~S10 are smaller than the ones shown in Figs.~S8 and S9, respectively. They are also nearly proportional to the coupling parameter $\eta$.

In the TPT calculations using the $\Delta$-ML models, the ensemble average in Eq.~(\ref{eq_15}) was taken over 1400 configurations selected randomly from 70-ps-NVT-ensemble FPMD simulations using the FP$_\mathrm{sl}$ method. Although these FPMD simulations are expensive, the overall computational time is still much smaller than full FP simulations. To ensure the applications of the second-order cumulant expansion, we show the probability distributions of the energy difference $\Delta U_{\kappa}^\mathrm{\Delta ML}$ in Fig.~S11. The distribution are well fitted by Gaussian functions, indicating that Eq.~(\ref{eq_15}) is reasonable approximations.

The MD simulations were performed using a Langevin thermostat~\cite{Allen_Book_2008}. For efficient sampling, the mass of hydrogen and time step were set to 2 amu and 1 fs. 

\subsection{MLFF and $\mathrm{\Delta}$-ML}

Similar to previous machine-learning approaches~\cite{Behler_PRL_2007, Bartok_PRL_2010}, the potential energy $U$ of a structure with $N_\mathrm{a}$ atoms in our MLFF method is approximated as a summation of local energies $U_{i}$ written as
\begin{align}
U &= \sum\limits_{i=1}^{N_\mathrm{a}} U_{i}. \label{eq_20}
\end{align}
Following the Gaussian approximation potential pioneered by B\'{a}rtok and co-workers~\cite{Bartok_PRL_2010}, the local energy $U_{i}$ is approximated as a weighted sum of functions $K\left( \mathbf{x}_{i}, \mathbf{x}_{i_\mathrm{B}} \right)$ 
centered at reference points $\{ \mathbf{x}_{i_\mathrm{B}} | i_\mathrm{B} =1,...,N_\mathrm{B} \}$
\begin{align}
U_{i} &= \sum\limits_{i_\mathrm{B}=1}^{N_\mathrm{B}} w_{i_\mathrm{B}} K\left(\mathbf{x}_{i},\mathbf{x}_{i_\mathrm{B}} \right). \label{eq_21}
\end{align}
The coefficients $\{ w_{i_\mathrm{B}} | i_\mathrm{B} =1,...,N_\mathrm{B} \}$ are optimized to best reproduce the FP energies, forces, and stress tensor components as obtained by the FPMD simulations. The descriptor $\mathbf{x}_{i}$ used in this study is a vector containing two and three body contributions~\cite{Jinnouchi_JCP_2020}:
\begin{align}
\mathbf{x}^{\mathrm{T}}_{i} \rightarrow \left( \sqrt{\beta^{(2)}} \mathbf{x}^{(2)\mathrm{T}}_{i}, \sqrt{\beta^{(3)}} \mathbf{x}^{(3)\mathrm{T}}_{i} \right), \label{eq_22}
\end{align}
Here, $\beta^{(2)}$ and $\beta^{(3)}(=1-\beta^{(2)})$ are the weights on the two and three body descriptors, $\mathbf{x}_{i}^{(2)}$ and $\mathbf{x}_{i}^{(3)}$, respectively. The vectors $\mathbf{x}_{i}^{(2)}$ and $\mathbf{x}_{i}^{(3)}$ collect the expansion coefficients of two and three body distribution functions with respect to the orthonormal radial and angular basis sets~\cite{Jinnouchi_PRB_2019, Jinnouchi_JCP_2020}:
\begin{align}
\rho_{i}^{(2)}\left( r \right) &= \frac{1}{\sqrt{4 \pi}} \sum\limits_{n=1}^{N^{0}_\mathrm{R}} c^{i}_{n} \chi_{n0}\left(r\right)  \label{eq_23} \\
\rho_{i}^{(3)}\left(r,s,\theta\right) &= \sum\limits_{l=0}^{L_\mathrm{max}} \sum\limits_{n=1}^{N^{l}_\mathrm{R}}\sum\limits_{\nu=1}^{N^{l}_\mathrm{R}} \sqrt{\frac{2l+1}{2}} \nonumber \\
&\times p_{n\nu l}^{i}\chi_{nl}\left(r\right)\chi_{\nu l}\left(s\right)P_{l}\left(\mathrm{cos}\theta\right). \label{eq_24}
\end{align}
The two and three body distribution functions $\rho_{i}^{(2)}$ and $\rho_{i}^{(3)}$ are defined as:
\begin{align}
\rho_{i}^{(2)}\left(r\right) &= \frac{1}{4\pi} \int \rho_{i} \left(r \hat{\mathbf{r}}\right)d\hat{\mathbf{r}}, \label{eq_25}  \\
\rho_{i}^{(3)}\left(r,s,\theta\right) &= \iint  d\hat{\mathbf{r}} d\hat{\mathbf{s}} \; \delta\left(\hat{\mathbf{r}}\cdot\hat{\mathbf{s}} - \mathrm{cos}\theta\right)  \nonumber \\
&\;\times  \rho_{i} \left( r\hat{\mathbf{r}} \right) \rho_{i} \left( s\hat{\mathbf{s}} \right),  \label{eq_26} \\
\rho_{i}\left(\mathbf{r}\right) &= \sum\limits_{j=1}^{N_{\mathrm{a}}} f_\mathrm{ cut} \left( \left| \mathbf{r}_{j} - \mathbf{r}_{i} \right| \right) g \left (\mathbf{r} - \left( \mathbf{r}_{j} - \mathbf{r}_{i} \right) \right)  \label{eq_27}
\end{align}
The function $g$ is the smoothed $\delta$-function, and $f_\mathrm{cut}$ is a cutoff function that smoothly eliminates the contribution from 
atoms outside a given cutoff radius $R_\mathrm{cut}$. For $\chi_{nl}$ and $P_{l}$, normalized spherical Bessel functions $\chi_{nl}= j_l(q_n r)$ and Legendre polynomials of order $l$ are used in this work, respectively.
For the kernel basis functions, the smooth overlap of atomic positions (SOAP) kernel~\cite{Bartok_PRB_2013} is employed
\begin{align}
K\left( \mathbf{x}_{i},\mathbf{x}_{i_\mathrm{B}} \right) &= \left( \mathbf{\hat{x}}_{i} \cdot \mathbf{\hat{x}}_{i_\mathrm{B}} \right)^{\zeta}. \label{eq_28}
\end{align}
The hat symbol $\mathbf{\hat{x}}_{i}$ denotes a normalized vector of $\mathbf{x}_{i}$. The normalization and exponentiation in Eq. (\ref{eq_28}) introduce non-linear terms that mix two- and three-body contributions.

The same formulation is used for the $\mathrm{\Delta}$-ML method. In the $\mathrm{\Delta}$-ML method, differences of potential energies and forces between two FP methods, semi-local and non-local functionals in this study, are used as the training data.

Parameter sets of the descriptors and kernel basis functions used in previous publications were employed in this study~\cite{Jinnouchi_PRB_2019, Jinnouchi_JCP_2020, Jinnouchi_JPCL_2023}. The parameters are tabulated in Table~S1.

Bulk solutions containing the redox species were modeled by systems as shown in Fig. 1. The number of water molecules was set to 32, 64, and 96. Three different model sizes were examined to clarify the system size effect. The sizes of the unit cells were set to obtain a water density of 0.99 g cm$^{-3}$. The size of the unit cell for the 32 water molecules is same as the one used in previous FPMD studies~\cite{Blumberger_JACS_2004, Blumberger_JCP_2006, Adriaanse_JCPL_2012,Liu_JPCB_2015}. For each of the reduced and oxidized states, MLFF and $\Delta$-ML models were constructed. All MLFF models were generated on the fly during a 100-ps-NVT-MD simulation at 400 K by using the active-learning algorithm developed in our previous study~\cite{Jinnouchi_PRB_2019}. The temperature for the training runs was set to a value higher than the target one of 298 K for production runs, to ensure that training data and kernel basis functions were provided in a wider phase space. A Langevin thermostat~\cite{Allen_Book_2008} was used to control the temperature. Exchange-correlation interactions between electrons were described by the semi-local RPBE functional~\cite{Hammer_PRB_1999} with Grimme's D3 dispersion corrections~\cite{Grimme_JCP_2010, Grimme_JCC_2011}. Probability distributions of the errors of the constructed MLFFs for energies and forces on test data are shown in Figs. S1 to S4 in SI. The RMSEs are similar to those of MLFFs used in previous studies~\cite{Jinnouchi_PRL_2019, Jinnouchi_PRB_2019, Jinnouchi_PRB_2020, Jinnouchi_JCP_2020}.

After examining the system size effect using the semi-local functional (see results in Fig. S13), $\Delta$-ML models were constructed on systems containing 64 water molecules. Each $\Delta$-ML model was trained on FP energies and forces of 40 structures selected randomly from a trajectory of a 20 ps NVT-ensemble FPMD simulation at 298K. The FPMD simulation was performed using the RPBE+D3 functional. Differences in energies and forces between the non-local and semi-local functionals for these 40 structures were used as training data. PBE0~\cite{Adamo_JCP_1999} with and without the Grimme's D3 dispersion correction~\cite{Grimme_JCP_2010, Grimme_JCC_2011} was employed as the non-local functional because the functional is known to accurately predict properties of water~\cite{Gaiduk_JCPL_2015}. The fraction of the exact exchange was set to 0.25 and 0.50 to determine how this influences the redox potentials. Error distributions of the $\Delta$-ML models on test structures are shown in Figs.~S2 to~S4. The RMSEs are one to two orders of magnitude smaller than the errors of the RPBE+D3 MLFFs.

The vacuum-water interface for the production run was modeled by a pure water slab without the redox species composed of 128 water molecules per unit cell. Following the previous study~\cite{Leung_JPCL_2010}, a rectangular cell of $12.5 \times 12.5 \times 50$ \AA $^3$ was employed. Similar to the MLFFs for the bulk solution systems containing the redox species, the MLFF for the interface was also generated by using the active-learning scheme. The systems used for the training were a pure water bulk composed of 64 water molecules in a $12.4 \times 12.4 \times 12.4$ \AA $^3$ cubic cell and pure water slab composed of 64 water molecules in an $8.8 \times 8.8 \times 40.8$ \AA $^3$ rectangular cell. Training simulations for both the bulk and slab were performed by NVT-ensemble MD simulations at 300, 400, 600 and 800 K. As shown in Fig.~S1, the constructed MLFF realizes small errors on test data taken from a 100-ps-MD simulation of a water slab composed of 128 water molecules at 298 K.

The annealing procesure used for the production runs explained in the previous subsection was also used to prepare for the initial structures for the training runs. All FP calculations were performed using VASP~\cite{Kresse_PRB_1996, Kresse_CMS_1996}. A 2$\times$2$\times$2 $\mathbf{k}$-point mesh was used for the bulk systems containing 32 water molecules. For other systems, $\Gamma$-point was used. Plane-wave cutoff energy was set to 520 eV. The PAW~\cite{Blochl_PRB_1994, Kresse_PRB_1999} distributed in VASP 5.4 was used in all FP calculations. The PAW atomic reference configuration was 1s$^{1}$ for H, 2s$^{2}$2p$^{4}$ for O, 3d$^{7}$4s$^{1}$ for Fe, and 4d$^{10}$5s$^{1}$ for Ag. The comparison of two atomic configurations for Cu, specifically 3d$^{10}$4p$^{1}$ and 3p$^{6}$3d$^{10}$4p$^{1}$, was conducted to examine the impact of semi-core electron relaxations on the redox potential. Upon verification that these effects are minimal within the PAW framework in VASP, as detailed in Section S7 in SI, we employed the less computationally demanding 3d$^{10}$4p$^{1}$ electronic configuration. The parameters for the MD simulations are same as the ones described in the previous subsection.

\section*{Acknowledgements}

We thank Carla Verdi for helpful discussions.

\section*{Code availability}

The VASP code is distributed by the VASP Software GmbH. The machine learning modules will be included in the release of vasp.6.3. Prerelease versions are available from G.K. upon reasonable request.

\section*{Data availability}

The data that support the findings of this study are available from the corresponding author upon reasonable request.

\nocite{*}
\bibliographystyle{naturemag}
\bibliography{./ms}
%\bibliography{aipsamp}% Produces the bibliography via BibTeX.

\end{document}